\documentclass[10pt]{article}

\usepackage{amsmath,amssymb}
\usepackage{graphicx}

\newcommand{\p}{\partial}

\begin{document}
\date{\today}
\title{Chimera states in heterogeneous networks}
\author{Carlo R. Laing\footnote{{\tt c.r.laing@massey.ac.nz}}\\
Institute of Information and Mathematical Sciences, \\ Massey University, Private Bag 102-904
NSMC, \\ Auckland, New Zealand}
\maketitle
\begin{abstract}
Chimera states in networks of coupled oscillators occur when some fraction of the oscillators synchronise
with one another, while the remaining oscillators are incoherent. Several groups have studied chimerae in networks
of identical oscillators, but here we study these states in 
heterogeneous models for which the natural frequencies
of the oscillators are chosen from a distribution. For a model consisting of two subnetworks we obtain 
exact results by reduction to a finite set of 
differential equations, and for a network of oscillators in a ring we generalise
known results. We find that heterogeneity can destroy chimerae, destroy all states except chimerae,
or destabilise chimerae in Hopf bifurcations, depending on the form of the heterogeneity.
\end{abstract}


{\bf Synchronisation of interacting oscillators is a problem of fundamental importance, with applications
from Josephson junction circuits to neuroscience~\cite{pikros01,str03,sin99,wiecol96}. Since oscillators are
unlikely to be identical, the effects of heterogeneity on their collective behaviour is of interest. One well-studied
system of heterogeneous phase oscillators is the Kuramoto model~\cite{acebon05,kuramoto,str00}, for which
there is global coupling. Generalisations of this model with nonlocal coupling~\cite{abrstr06,setsen08,abrstr04,kurbat02,omemai08}, 
or several populations of oscillators~\cite{abrmir08},
have shown interesting types of behaviour referred to as ``chimera'' states
in which some oscillators are synchronised with one another while the remainder are incoherent.
Even though the effects of heterogeneity on synchronisation have been emphasised in the
past~\cite{acebon05,laikev08,kuramoto,str00}, all chimerae have so far been studied in networks of
identical oscillators. This raises the obvious question: do chimerae exist in
networks of nonidentical oscillators? Here we address
the question analytically, first using recent results to exactly derive a finite set of differential equations governing the
dynamics of chimerae in two coupled networks of heterogeneous phase oscillators, and then using a similar idea
to extend the results of Abrams and Strogatz~\cite{abrstr06} for chimera states in a ring of coupled oscillators.}

\section{Introduction}

Networks of coupled oscillators have been studied for many 
years~\cite{str03,laikev08,abrmir08,abrstr06,pikros01,renerm00,baeguc91}. 
One well-known system is the Kuramoto
model~\cite{acebon05,kuramoto,barhun08,kurbat02,str00} of phase oscillators. In the last few years several authors have 
studied ``chimera'' states in networks of {\it identical} Kuramoto
oscillators, in which some oscillators are synchronised with one another while the remainder are 
incoherent. Much analytical progress has been made
in the study of these states~\cite{abrmir08,abrstr06,setsen08,abrstr04,kurbat02,omemai08}.
It is very unlikely that any physical system being modelled by a network of coupled oscillators will have identical
units, so the robustness of chimerae to network heterogeneity is naturally of interest. Certain networks
of coupled oscillators are known to have non-generic properties~\cite{watstr93,watstr94}, and it is of interest to know whether chimera
states are generic and stable (and thus expected to be generally observed) or not.

Here we conduct an analytical investigation into the robustness of pre\-vi\-ous\-ly-studied chimerae with respect to heterogeneity in the
intrinsic frequencies of oscillators. We find that chimerae are robust with respect to this type of heterogeneity,
and  show some of the bifurcations that chimera and other states undergo as the oscillators in the network
are made more dissimilar. Our results provide more evidence that the Ott-Antonsen ansatz~\cite{ottant08}
correctly describes attracting states in Kuramoto-type networks when the oscillators are not identical.
 Some of the ideas here have recently been used by others to study
a single population of oscillators with a bimodal frequency distribution~\cite{marbar08} and the periodically forced 
Kuramoto model~\cite{chistr08}. 

Note that the term ``chimera'' has been used in the past to refer to certain states in networks of identical 
oscillators~\cite{omemai08,setsen08,abrmir08,abrstr06}, but here we also use the term to describe similar states
in heterogeneous networks in the obvious way. A state found by numerically continuing from a chimera state in a network
of identical oscillators is also referred to as a chimera.

In Sec.~\ref{sec:model} we present the first model of two coupled networks, then consider its continuum limit
and use the remarkable recent result of
Ott and Antonsen~\cite{ottant08} to derive three ODEs which exactly describe some of its behaviour. 
In Sec.~\ref{sec:results} we perform a
limited bifurcation analysis
of these ODEs and interpret the results. In Secs.~\ref{sec:other} and~\ref{sec:gen} we consider 
other distributions of the intrinsic oscillator frequencies, and generalisations, respectively.
In Sec.~\ref{sec:ring} we consider oscillators on a ring.

\section{Two coupled networks }\label{sec:model}
We first consider two networks of coupled oscillators with uniform coupling between oscillators within each network, and a weaker
coupling to those in the other network.
Our model equations are
\begin{equation}
   \frac{d\theta_i^k}{dt}  =  \omega_i^k+\sum_{m=1}^{2}\frac{K_{km}}{N}\sum_{j=1}^N\sin{(\theta_j^m-\theta_i^k-\alpha)} \label{eq:dtheta} 
\end{equation}
for $i=1,\ldots N$ and $k=1,2$, where the natural frequencies $\omega_i^k$ are chosen from a distribution $g_k(\omega^k)$. 
Our system is the same as that of Montbri\'o et al.~\cite{monkur04};
the system of Abrams et al.~\cite{abrmir08} is a special case of that studied here. A similar system was studied by Barreto et al.~\cite{barhun08},
but their focus was the onset of synchrony, as was Montbri\'o et al's. 
Like Abrams et al.~\cite{abrmir08}, we choose $K_{11}=K_{22}=\mu$
and $K_{12}=K_{21}=\nu$, set $\mu+\nu=1$ (by rescaling time if necessary), and choose $\mu>\nu$. We define $A=\mu-\nu$ and
$\beta=\pi/2-\alpha$. Abrams et al. found that for $\beta$ and $A$ sufficiently small and positive (and all
$\omega_i^k$ equal), both the
completely synchronised state ($\theta_j^m=\theta_i^k$ for all $j,i,m,k$) and the chimera state (all
oscillators in one population perfectly synchronised, all oscillators in the other population incoherent)
were stable.

We take the continuum limit of~(\ref{eq:dtheta}), letting $N\rightarrow\infty$. 
The system is then described by the probability density function (PDF) $f_k(\omega^k,\theta^k,t)$
for each population $k$. We define two order parameters
\begin{equation}
    z_{k}(t)=\int_{-\infty}^{\infty}\int_0^{2\pi}\exp{(i\theta^k)}f_{k}(\omega^k,\theta^k,t)\ d\theta^k\  d\omega^k \label{eq:ztheta}
\end{equation}
for $k=1,2$.
Each $f_k$ satisfies a continuity equation
\begin{equation}
    \frac{\p f_k}{\p t}+\frac{\p}{\p \theta^k}(f_kv_k)=0 \label{eq:feq}
\end{equation}
where the velocity
\begin{eqnarray}
   v_k & = & \omega^k+\{\exp{[-i(\theta^k+\alpha)]}(\mu z_{k}+\nu z_{k'}) \nonumber \\
   & & -\exp{[i(\theta^k+\alpha)]}(\mu \bar{z}_{k}+\nu \bar{z}_{k'})\}/(2i),
\end{eqnarray}
$k'=3-k$, and
an overbar denotes the complex conjugate.
Writing $f_k$ as a Fourier series in $\theta^k$ we have
\begin{equation}
    f_k(\omega^k,\theta^k,t)=\frac{g_k(\omega^k)}{2\pi}\left[1+\left\{\sum_{n=1}^{\infty}h_n(\omega^k,t)\exp{(in\theta^k)}+c.c.\right\}\right] 
\label{eq:fourier}
\end{equation}
where ``$c.c.$'' denotes the complex conjugate of the previous term.
Substituting~(\ref{eq:fourier}) into~(\ref{eq:feq}) and~(\ref{eq:ztheta}),
one can derive an infinite set of integro-differential equations for the $h_n$~\cite{monkur04}.
However, Ott and Antonsen~\cite{ottant08} noticed that for the special choice
\begin{equation}
    h_n(\omega^k,t)=\left[a_k(\omega^k,t)\right]^n \label{eq:ans}
\end{equation}
i.e.~$h_n$ is $a_k$ raised to the $n$th power,
all of these differential equations are actually the same, and we are left with a single PDE
governing the dynamics of $a_k(\omega^k,t)$:
\begin{eqnarray}
    \frac{\p a_k}{\p t}+i\omega^k a_k-(e^{i\alpha}/2)(\mu\bar{z}_k+\nu\bar{z}_{k'}) & & \nonumber \\
 +(e^{-i\alpha}/2)(\mu z_k+\nu z_{k'})a_k^2 & = & 0
\label{eq:atheta}
\end{eqnarray}
where
\begin{equation}
   z_k(t)=\int_{-\infty}^{\infty}\bar{a}_k(\omega^k,t)g_k(\omega^k)\ d\omega^k \label{eq:zth}
\end{equation}
The ansatz~(\ref{eq:ans}) is not trivial, and the reduction from an infinite set of differential equations to
one is remarkable. 
Ott and Antonsen~\cite{ottant08} give more detail on the circumstances under which this ansatz is valid, and we
discuss its usefulness in describing {\it attracting} states below.

As is well-known~\cite{monkur04,ottant08,kuramoto,marbar08}, if $g_k$ 
is a Lorentzian distribution the integral~(\ref{eq:zth}) can be evaluated analytically.
Suppose that
\begin{eqnarray}
    g_k(\omega^k) & = & \frac{D_k/\pi}{(\omega^k-\Omega_k)^2+D_k^2} 
\end{eqnarray}
i.e.~the $\omega^k_i$ are from a distribution centred at $\Omega_k$ with half-width-at-half-maximum $D_k$.
Then $z_k(t)=\bar{a}_k(\Omega_k-iD_k,t)$ and
evaluating~(\ref{eq:atheta}) at $\omega^k=\Omega_k-iD_k$ we obtain
\begin{eqnarray}
    \frac{d\bar{z}_k}{\p t}+(D_k+i\Omega_k) \bar{z}_k-(e^{i\alpha}/2)(\mu\bar{z}_k+\nu\bar{z}_{k'}) & & \nonumber \\
   +(e^{-i\alpha}/2)(\mu z_k+\nu z_{k'})\bar{z}_k^2 & = & 0
\label{eq:dbarzps}
\end{eqnarray}
i.e.~a complex ODE for each $k$. Writing $z_1=r_1e^{-i\phi_1}$ and $z_2=r_2e^{-i\phi_2}$ 
and defining $\phi=\phi_1-\phi_2$ we obtain the three real ODEs:
\begin{eqnarray}
    \frac{dr_1}{dt} & = & -D_1 r_1+\left(\frac{1-r_1^2}{2}\right)[\mu r_1\cos{\alpha}+\nu r_2\cos{(\phi-\alpha)}] \label{eq:dr1} \\
    \frac{dr_2}{dt} & = & -D_2 r_2+\left(\frac{1-r_2^2}{2}\right)[\mu r_2\cos{\alpha}+\nu r_1\cos{(\phi+\alpha)}] \label{eq:dr2} \\
   \frac{d\phi}{dt} & = & \left(\frac{r_1^2+1}{2r_1}\right)[\mu r_1\sin{\alpha}-\nu r_2\sin{(\phi-\alpha)}]+\Omega_2 \nonumber \\
& - & \left(\frac{r_2^2+1}{2r_2}\right)[\mu r_2\sin{\alpha}+\nu r_1\sin{(\phi+\alpha)}] \label{eq:dphi}
\end{eqnarray}
where, without loss of generality, we have set $\Omega_1=0$.
When $D_1=D_2=\Omega_2=0$, we recover the results of Abrams et al.~\cite{abrmir08}. In particular, $r_1=1$ ($\theta^1_i$ all equal) 
is invariant. If $r_1=1$, there also exists the perfect synchrony state ($r_2=1,\phi=0$) and, depending 
on parameters (see Fig.~4
in~\cite{abrmir08}) two other fixed points with $r_2\neq 1$ (the chimerae). 
When they exist, one of these fixed points is a saddle while the other
is either stable or unstable, changing stability via a supercritical Hopf bifurcation. We now proceed with a limited analysis
of~(\ref{eq:dr1})-(\ref{eq:dphi}).

\section{Results}\label{sec:results}
\subsection{Varying distribution widths, no frequency offset}
\begin{figure}
\begin{center}
\includegraphics[width=12cm]{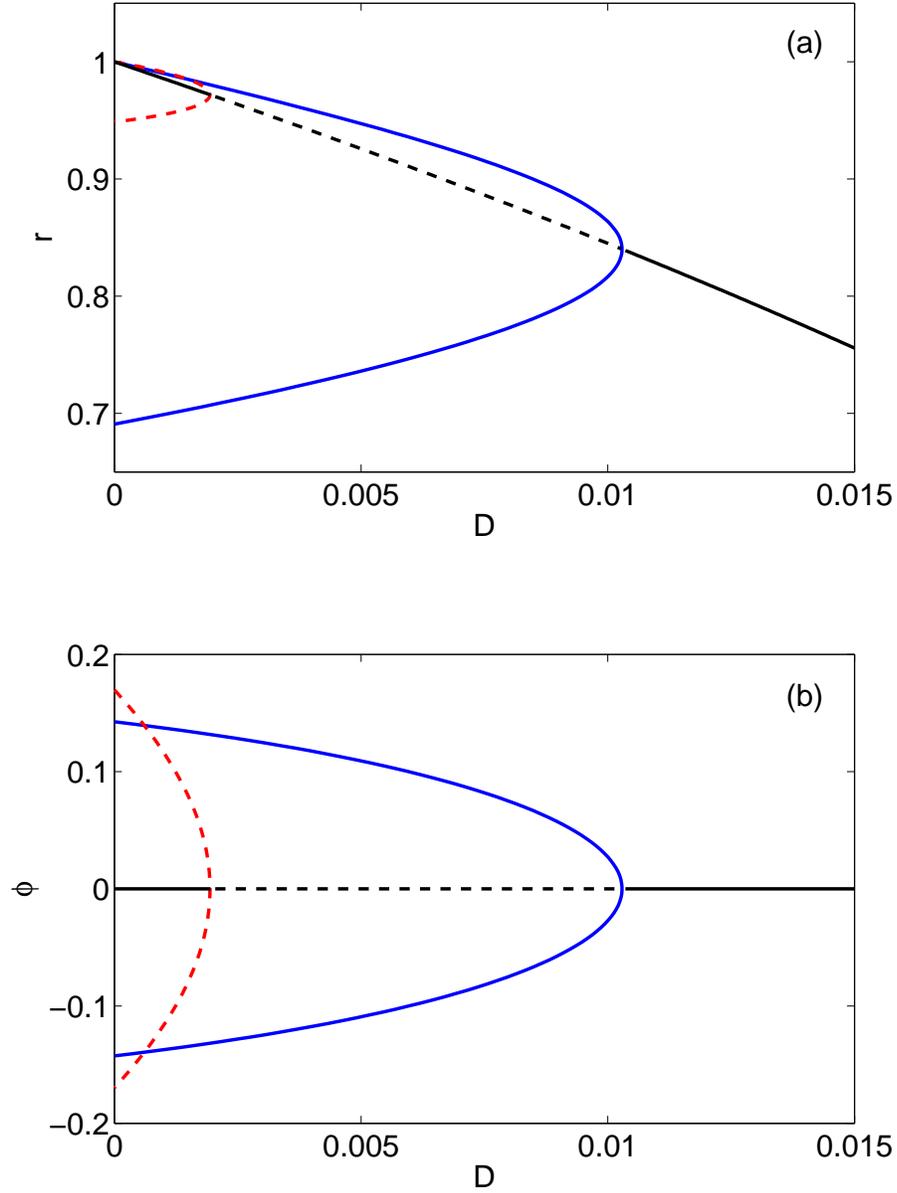}
\caption{(Colour online) Fixed points of~(\ref{eq:dr1})-(\ref{eq:dphi}) when $D_1=D_2=D, \Omega_2=0$. (a): $r_1$ and $r_2$ as a function of $D$. 
(b): $\phi$ as a function of $D$.
Blue curve: stable chimera. Red curve: saddle chimera. Black curve: symmetric state ($r_1=r_2$).
Solid lines indicate stable solutions, dashed lines unstable.
Other parameters: $A=0.2,\beta=0.07$.}
\label{fig:symm}
\end{center}
\end{figure}

First consider the case when $D_1=D_2=D$ and $\Omega_2=0$. The system~(\ref{eq:dr1})-(\ref{eq:dphi})
 possesses $\mathbb{Z}_2$ symmetry:
$(r_1,r_2,\phi)\rightarrow(r_2,r_1,-\phi)$. We choose $A=0.2$ and $\beta=0.07$ so that for $D=0$, there exist
five fixed points of~(\ref{eq:dr1})-(\ref{eq:dphi}): the perfect synchrony state $(r_1,r_2,\phi)=(1,1,0)$ and 
two chimerae (one stable and one a saddle) for which $r_1=1$ and $r_2\neq 1$ and $\phi\neq 0$,
and their symmetrically related states. The fixed points of~(\ref{eq:dr1})-(\ref{eq:dphi}) and their stability 
as a function of $D$ 
 are shown in Fig.~\ref{fig:symm}.
There are several interesting observations to be made here. Firstly (for these parameter values), 
increasing the heterogeneity 
of the network first destabilises the symmetric state ($r_1=r_2$), then restabilises it. Secondly, increasing
the heterogeneity actually {\it decreases}
 the width of the angular distribution of the unsynchronised population
in the chimera state (lower blue branch in panel (a) of Fig.~\ref{fig:symm}).

\subsection{Varying one distribution width, no frequency offset}
\begin{figure}
\begin{center}
\includegraphics[width=12cm]{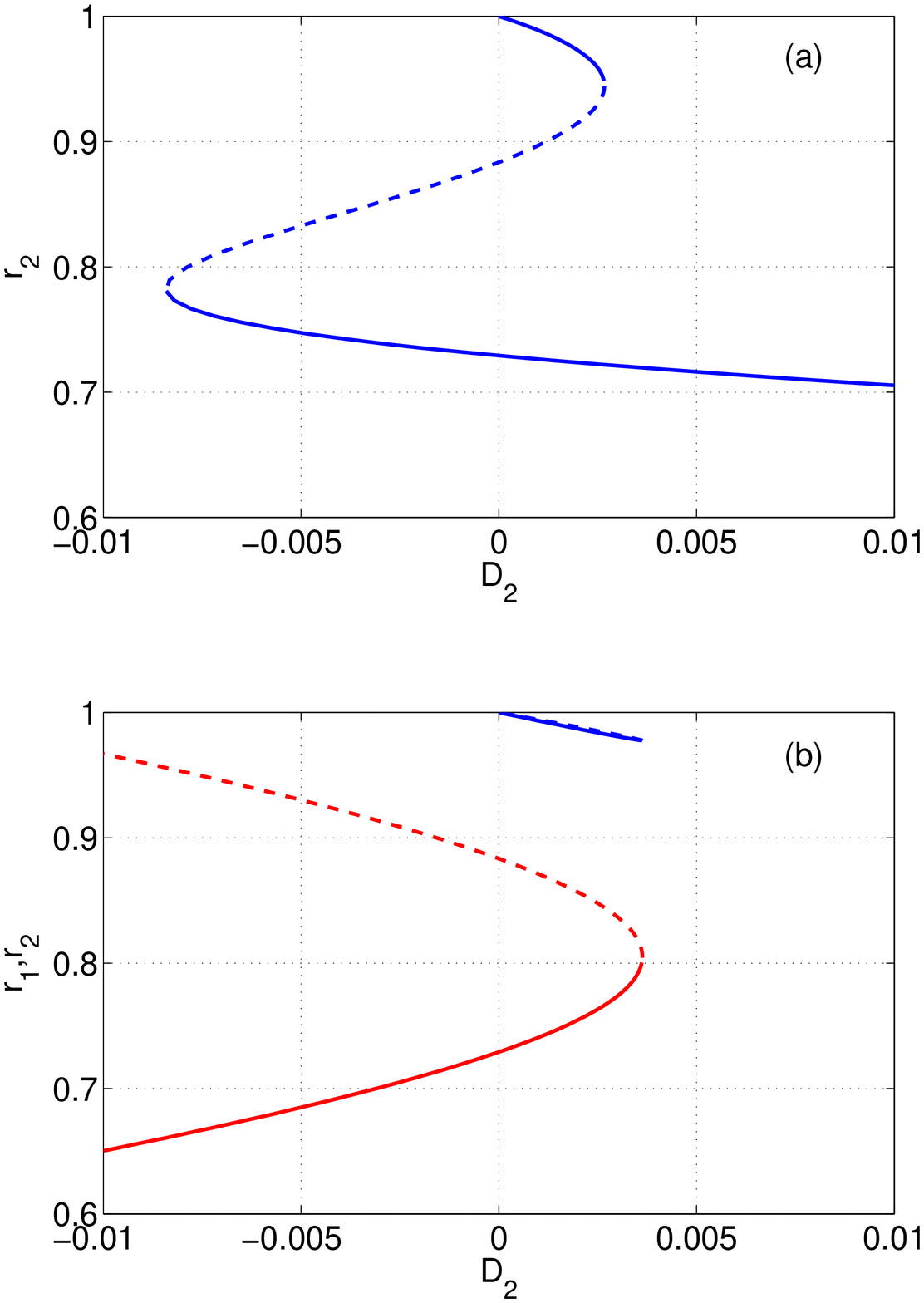}
\caption{(Colour online) Fixed points of~(\ref{eq:dr1})-(\ref{eq:dphi}) when $D_1=\Omega_2=0$. (a): $r_2$ as a function of $D_2$ when $r_1=1$. 
(b): $r_1$ (red) and $r_2$ (blue) as a function of $D_2$.
Solid lines indicate stable solutions, dashed lines unstable.
Other parameters: $A=0.2,\beta=0.1$.}
\label{fig:varyD}
\end{center}
\end{figure}

Now consider varying $D_2$, while $D_1=\Omega_2=0$. The system no longer possesses any symmetry,
so the effect of increasing $D_2$ from zero on the chimerae with $r_1=1,r_2\neq 1$ will be different from its
effect on the chimerae with $r_1\neq 0,r_2=1$. We choose $A=0.2$ and $\beta=0.1$, so that as before,
when $D_2=0$ there exist five fixed points of~(\ref{eq:dr1})-(\ref{eq:dphi}). Results are shown in Fig.~\ref{fig:varyD}.
With five solutions to track, we do not show $\phi$. Also, even though $D_2<0$ is not physically meaningful
we plot fixed points for $D_2<0$ to show how branches of solutions are connected.

Panel (a) in Fig.~\ref{fig:varyD} shows fixed points for which $r_1=1$ (recall that we are making population 2 heterogeneous).
As $D_2$ is increased, the perfectly synchronous solution that exists at $D_2=0$ is destroyed in a saddle-node
bifurcation, while the chimera with population 2 desynchronised persists. Panel (b)
shows the fate of the two chimerae (one stable and one a saddle) for which $r_2=1$ when $D_2=0$.
We see that they are both destroyed in a saddle-node bifurcation as $D_2$ is  increased. From this figure 
we see that if one population is made sufficiently heterogeneous, the only solution that persists is the chimera
for which the oscillators in that population are desynchronised.
Interestingly, if $D_2$ is increased to larger values ($D_2\approx 0.1$), the state where
both populations are in the ``splay'' state, with uniform angular density, i.e.~$r_1=r_2=0$ and $\phi$ is no longer
meaningful, becomes stable (not shown). 

\subsection{Varying frequency offset $\Omega_2$}
\begin{figure}
\begin{center}
\includegraphics[width=10cm]{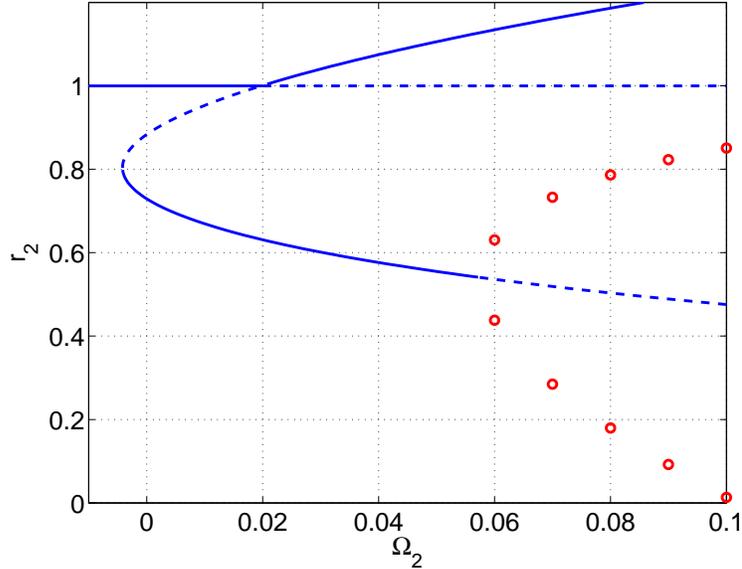}
\caption{(Colour online) $r_2$ as a function of $\Omega_2$, when $r_1=1$.
Solid lines indicate stable fixed points of~(\ref{eq:dr1})-(\ref{eq:dphi}), dashed lines unstable. 
Circles are the maximum and minimum
of $r_2$ during stable periodic oscillations. Note that the branch with $r_2>1$ is not physically
meaningful.
Other parameters: $A=0.2,\beta=0.1$.}
\label{fig:varyom}
\end{center}
\end{figure}

We now consider varying $\Omega_2$ with $D_1=D_2=0$. 
Since we are interested in states for which at least one of the
populations is in complete synchrony we set $r_1=1$ and only consider~(\ref{eq:dr2})-(\ref{eq:dphi}).
The completely synchronised state, $(r_2,\phi)=(1,0)$, exists when $\Omega_2=0$, and as $\Omega_2$
is increased it persists as the fixed point $(r_2,\phi)=(1,\phi)$, where $\phi$ is the solution closest
to zero of $2\nu\cos{\alpha}\sin{\phi}=\Omega_2$. Numerical results
are shown in Fig.~\ref{fig:varyom}.
We see that as $\Omega_2$ is increased from zero, the synchronised state for which $r_2=1$ is destroyed in a 
transcritical bifurcation involving the saddle chimera, while the stable chimera undergoes a supercritical
Hopf bifurcation, leading to oscillations  in $r_2$ and $\phi$. However, decreasing $\Omega_2$ from zero
causes destruction of the stable chimera in a saddle-node bifurcation with the saddle chimera. For these parameter
values, one can see that the stable chimera is much more robust to speeding up the asynchronous oscillators,
as opposed to a slowing them down.

\subsection{Discussion}
Our bifurcation analysis in this section has found all four codimension-1 bifurcations of ODEs (saddle-node, pitchfork,
transcritical and Hopf) and a two-parameter study is likely to find higher codimension bifurcations.

Pikovsky and Rosenblum~\cite{pikros08} recently studied~(\ref{eq:dtheta}) with identical $\omega_i^k$ (i.e.~the system of
Abrams et al.~\cite{abrmir08}, and our system when $D_1=D_2=\Omega_2=0$)  as a special case
and found that the ansatz~(\ref{eq:ans}) did not completely describe the
possible dynamics of this system. However, Pikovsky and Rosenblum~\cite{pikros08} and
other authors~\cite{marbar08} found that when the oscillators are non-identical,
this ansatz does successfully allow one to describe attracting states. We also find this behaviour here: 
extensive numerical simulations show that the stable states
shown in Figs.~\ref{fig:symm} and~\ref{fig:varyD} for $D,D_2>0$ are attracting, and that the angular
distributions of these stable states are given by~(\ref{eq:fourier})-(\ref{eq:ans}), even if the initial
distributions are not. However, the same cannot be said for the results in Fig.~\ref{fig:varyom}, for which
oscillators within each of the two networks are identical.
This figure correctly predicts the dynamics if the initial angular distribution is given 
by~(\ref{eq:fourier})-(\ref{eq:ans}) but other initial conditions give solutions not described by 
Fig.~\ref{fig:varyom} (not shown). This relationship between initial conditions and dynamics when oscillators
within each network are identical was also
noticed by Montbri\'o et al.~\cite{monkur04}.
The results in Fig.~\ref{fig:varyom} are likely to be a subset of those that could be found using the
approach in Ref.~\cite{pikros08}.

In related results, 
Montbri\'o et al.~\cite{monkur04} fixed $\Omega_2\neq 0$ and varied both $\alpha$ and $\nu$ and found chimera states,
both for homogeneous and heterogeneous networks.



\section{Other distributions}\label{sec:other}
Now we consider the effects of choosing the $\omega^k_i$ from distributions other than the Lorentzian, first
numerically and then analytically.
\subsection{Gaussian distribution: numerical simulations}
\begin{figure}
\begin{center}
\includegraphics[width=10cm]{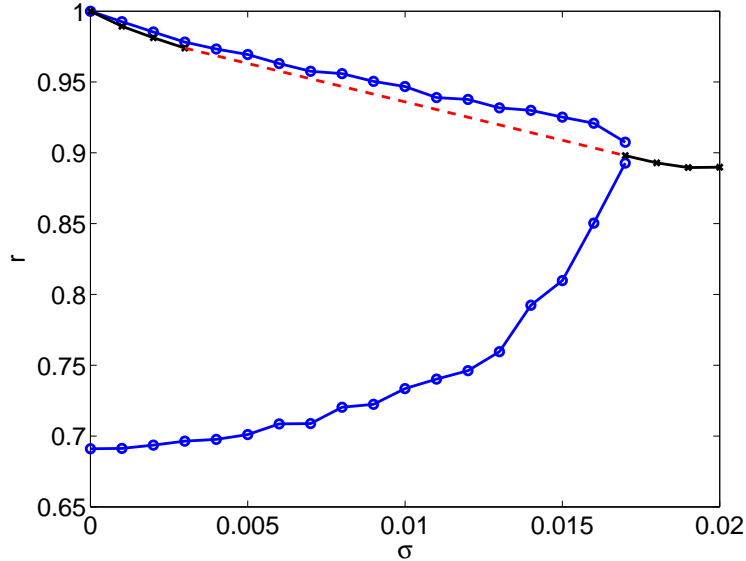}
\caption{(Colour online) $r_1$ and $r_2$ fitted to simulations of~(\ref{eq:dtheta}),
where all $\omega_i^k$  are chosen from a normal distribution of mean zero and 
standard deviation $\sigma$. Blue circles joined by a line: stable chimera.
Black crosses joined by a line: stable symmetric state ($r_1=r_2$). Red dashed line: presumed unstable
symmetric state. Compare with Fig.~\ref{fig:symm}. See the text for details on the fitting.
Other parameters: $A=0.2,\beta=0.07, N=1000$.}
\label{fig:gauss}
\end{center}
\end{figure}

Figure~\ref{fig:gauss} shows the results of fitting the time-dependent PDF
\begin{eqnarray}
    f_k(\theta,t) & = & \frac{1}{2\pi}\left[1+\left\{\sum_{n=1}^{\infty}\left(r_ke^{i\phi_k}\right)^ne^{in\theta}+c.c.\right\}\right] \nonumber \\
   & = & \frac{1-r_k^2}{2\pi[1-2r_k\cos{(\phi_k-\theta)}+r_k^2]} \label{eq:PDF}
\end{eqnarray}
to each population in simulations 
of~(\ref{eq:dtheta}) after transients, where all $\omega_i^k$ are chosen from a
normal distribution of mean zero and standard deviation $\sigma$. We found that both $r_1$ and $r_2$ tended to
constant values, as did $\phi_2-\phi_1$ (not shown). Only stable states are shown in Fig.~\ref{fig:gauss},
but the results are compatible with those shown in Fig.~\ref{fig:symm},
suggesting that there is nothing special about the Lorentzian distribution, as has been
noted by others~\cite{marbar08,chistr08}. The unstable states could
presumably be found using the ``equation-free'' method~\cite{laikev08,kevgea04,lai06} of analysing low-dimensional
descriptions of high-dimensional systems, under the assumption that these states
are also exactly described by the variables $r,\phi$ for each population.

\subsection{Another distribution} \label{sec:another}
As an alternative~\cite{ottant08}, we 
suppose that the $\omega_i^k$ are chosen from the distribution
\begin{equation}
    g_k(\omega)=\frac{\sqrt{2}s_k^3}{\pi}\left(\frac{1}{\omega^4+s_k^4}\right)
\end{equation}
which has mean zero (for simplicity) and variance $s_k^2$.
This $g_k(\omega)$ has poles at $\omega =s_k(\pm 1\pm i)/\sqrt{2}$, and the integral~(\ref{eq:zth})
gives
\begin{equation}
   z_k=\left(\frac{1+i}{2}\right)z_{k}^{-}+\left(\frac{1-i}{2}\right)z_{k}^{+}
\end{equation}
where $z_k^{\pm}$ satisfy
\begin{eqnarray}
    \frac{dz^\pm_k}{dt}-\left[s_k(1\pm i)/\sqrt{2}\right]z^\pm_k-(e^{-i\alpha}/2)(\mu z_k+\nu z_{k'})  \nonumber \\
+(e^{i\alpha}/2)(\mu\bar{z}_k+\nu\bar{z}_{k'})\left(z^\pm_k\right)^2  =  0
\end{eqnarray}
Thus we have four coupled complex ODEs, rather than the two~(\ref{eq:dbarzps}).
This can clearly be generalised to other distributions which are rational functions of $\omega$.


\section{Generalisations}\label{sec:gen}
We now briefly mention several generalisations of the results above.
Suppose that the system~(\ref{eq:dtheta}) is periodically forced, i.e.~we add the term 
$\Lambda_k\sin{(\widehat{\Omega}t-\theta_i^k)}$ to~(\ref{eq:dtheta}),
as done recently for a single population~\cite{chistr08}.
Going to a
coordinate frame rotating with angular frequency $\widehat{\Omega}$, we find 
that~(\ref{eq:dbarzps}) is replaced by
\begin{eqnarray*}
     \frac{d\bar{z}_k}{\p t}+(D_k+i(\Omega_k-\widehat{\Omega})) \bar{z}_k-\left[\frac{\Lambda_k+e^{i\alpha}(\mu\bar{z}_k+\nu\bar{z}_{k'})}{2}\right] \nonumber \\
  +\left[\frac{\Lambda_k+e^{-i\alpha}(\mu z_k+\nu z_{k'})}{2}\right]\bar{z}_{k}^2=0
\end{eqnarray*}
The system is no longer invariant under a translation of time, so during the derivation of ODEs 
like~(\ref{eq:dr1})-(\ref{eq:dphi}) we find that 
we need both $\phi_1$ and $\phi_2$, not just their difference. 

Another possibility is that there is a uniform delay of $\tau$ between the two populations, but zero delay within them,
i.e.~we replace $\theta_j^m$ in~(\ref{eq:dtheta}) by $\theta_j^m(t-\tau)$ when $m=k'$. The effect of this
is to replace $z_{k'}(t)$ in~(\ref{eq:dbarzps}) by $z_{k'}(t-\tau)$ and 
the equivalent of~(\ref{eq:dr1})-(\ref{eq:dphi}) is now the four delay differential equations:
\begin{eqnarray*}
   \frac{dr_1}{dt}+D_1 r_1+\left(\frac{r_1^2-1}{2}\right)[\mu r_1\cos{\alpha}+\nu r_2(t-\tau)\cos{\{\phi_1-\phi_2(t-\tau)-\alpha\}}] & = & 0  \\
    \frac{d\phi_1}{dt}-\left(\frac{r_1^2+1}{2r_1}\right)[\mu r_1\sin{\alpha}-\nu r_2(t-\tau)\sin{\{\phi_1-\phi_2(t-\tau)-\alpha\}}] & = & 0  \\
    \frac{dr_2}{dt}+D_2 r_2+\left(\frac{r_2^2-1}{2}\right)[\mu r_2\cos{\alpha}+\nu r_1(t-\tau)\cos{\{\phi_2-\phi_1(t-\tau)-\alpha\}}] & = & 0  \\
    \frac{d\phi_2}{dt}+\Omega_0-\left(\frac{r_2^2+1}{2r_2}\right)[\mu r_2\sin{\alpha}-\nu r_1(t-\tau)\sin{\{\phi_2-\phi_1(t-\tau)-\alpha\}}] & = & 0
\end{eqnarray*}
If all of the variables on the RHS of~(\ref{eq:dtheta}) were delayed by $\tau$, we could define
$\phi=\phi_1-\phi_2$ as before, and derive three coupled DDEs rather than four above. The analysis of the 
equations in this section remains an open problem.

\section{Oscillators on a ring}\label{sec:ring}
We now consider a ring of oscillators, with non-local coupling between them.
The original presentation of chimerae was in such a system, with identical
oscillators~\cite{abrstr06,abrstr04,kurbat02}. The chimera state for this system consists of oscillators 
on one part of the ring being synchronised,
while over the remainder of the ring they are incoherent.  Simulations reported in~\cite{abrstr06}
indicate that these states are also robust with respect to perturbations of the oscillators' natural frequencies,
and we now show how to use the ideas above to investigate this analytically.

Consider the model consisting of oscillators on a ring studied in
refs.~\cite{kurbat02,abrstr06}, but include heterogeneity in the intrinsic frequencies
of the oscillators.
The system is
\begin{equation}
   \frac{d\phi_i}{dt}  =  \omega_i-\frac{2\pi}{N}\sum_{j=1}^NG\left(\frac{2\pi|i-j|}{N}\right)\cos{(\phi_i-\phi_j-\beta)} \label{eq:nettheta} 
\end{equation}
for $i=1,\ldots N$, where the natural frequencies $\omega_i$ are chosen from a distribution $g(\omega)$. 
The coupling function $G$ is periodic with period $2\pi$. Equation~(\ref{eq:nettheta}) is the discrete version of
\begin{equation}
   \frac{\p \phi}{\p t}=\omega-\int_0^{2\pi}G(x-y)\cos{[\phi(x,t)-\phi(y,t)-\beta]}dy \label{eq:dphinet}
\end{equation}
which for constant $\omega$ is the same as that studied by~\cite{kurbat02,abrstr06}. The analysis
for a heterogeneous network is very similar to that for a network of identical oscillators, so we skip many of the
details here and refer the reader to~\cite{abrstr06}. The main difference is that $\omega$ is now a variable,
and certain quantities now have to be replaced by integrals over $\omega$, weighted by $g(\omega)$.

\subsection{Analysis}
First we go to a rotating reference frame with angular speed $\Omega$, i.e.~let $\theta=\phi-\Omega t$.
Then we define an order parameter
\[
   R(x,t)e^{i\Theta(x,t)}=\int_0^{2\pi}G(x-y)e^{i\theta(y,t)} dy
\]
so that~(\ref{eq:dphinet}) can be written
\begin{equation}
    \frac{\p \theta}{\p t}=\omega-\Omega-R\cos{(\theta-\Theta-\beta)} \label{eq:ptheta}
\end{equation}
We now look for stationary states, so that $R$ and $\Theta$ are independent of $t$.
At position $y$, if $R(y)>|\omega-\Omega|$, then the oscillators will move to the stable fixed point
$\theta^{\ast}$, which is given by the solution of
\[
   \omega-\Omega=R\cos{[\theta^{\ast}-\Theta-\beta]}
\]
For those drifting oscillators at $y$ with $R(y)<|\omega-\Omega|$, we replace $e^{i\theta(y)}$ in the order parameter
definition with its average over $\theta$~\cite{abrstr06,kurbat02}, but now weighted by $g(\omega)$ (over the appropriate range of $\omega$).
After some calculation the result is that at stationarity we have
\[
    R(x)e^{i\Theta(x)}=
\]
\begin{equation}
    e^{i\beta}\int_0^{2\pi}G(x-y)e^{i\Theta(y)}\int_{-\infty}^{\infty}\left(\frac{\omega-\Omega-\sqrt{(\omega-\Omega)^2-R^2(y)}}{R(y)}\right) g(\omega)d\omega\ dy \label{eq:order}
\end{equation}
In some cases this double integral can be exactly evaluated. We follow~\cite{abrstr06} and suppose that $G(x)=(1+A\cos{x})/(2\pi)$,
so that $G(x-y)=(1+A\cos{x}\cos{y}+A\sin{x}\sin{y})/(2\pi)$. Let us define
\[
   h(y)=\int_{-\infty}^{\infty}\left(\frac{\omega-\Omega-\sqrt{(\omega-\Omega)^2-R^2(y)}}{R(y)}\right) g(\omega)d\omega
\]
Thus under the assumption that $R$ and $\Theta$ are even (which can be shown to be self-consistent)
\begin{equation}
    R(x)e^{i\Theta(x)}=c+a\cos{x} \label{eq:Rtrig}
\end{equation}
where
\begin{equation}
    c=\frac{e^{i\beta}}{2\pi}\int_0^{2\pi} e^{i\Theta(y)}h(y) dy \label{eq:c2}
\end{equation}
and
\begin{equation}
   a=\frac{Ae^{i\beta}}{2\pi}\int_0^{2\pi} e^{i\Theta(y)}h(y)\cos{y}\  dy \label{eq:a2}
\end{equation}
Since~(\ref{eq:order}) is unchanged by the shift $\Theta(x)\rightarrow\Theta(x)+\Theta_0$, we can take $c$ to be real. 
To write the right hand sides of~(\ref{eq:c2}) and~(\ref{eq:a2})  in terms of $a$ and $c$, note that
\[
    R^2(y)=\left[R(y)e^{i\Theta(y)}\right]\left[R(y)e^{-i\Theta(y)}\right]=c^2+2c\mbox{Re}(a)\cos{y}+|a|^2\cos^2{y}
\]
and
\begin{eqnarray*}
    e^{i\Theta(y)}h(y) & = & R(y)e^{i\Theta(y)}\frac{h(y)}{R(y)} \\
   & = & (c+a\cos{y})\int_{-\infty}^{\infty}\left(\frac{\omega-\Omega-\sqrt{(\omega-\Omega)^2-R^2(y)}}{R^2(y)}\right) g(\omega)d\omega \\
   & = & \frac{1}{c+\bar{a}\cos{y}}\int_{-\infty}^{\infty}\left(\omega-\Omega-\sqrt{(\omega-\Omega)^2-R^2(y)}\right) g(\omega)d\omega
\end{eqnarray*}
where overbar indicates complex conjugate. Thus we have
\begin{equation}
    c=\frac{e^{i\beta}}{2\pi}\int_0^{2\pi} \frac{1}{c+\bar{a}\cos{y}}\int_{-\infty}^{\infty}f(\omega,y) g(\omega)d\omega\  dy \label{eq:c}
\end{equation}
and
\begin{equation}
   a=\frac{Ae^{i\beta}}{2\pi}\int_0^{2\pi} \frac{\cos{y}}{c+\bar{a}\cos{y}}\int_{-\infty}^{\infty}f(\omega,y) g(\omega)d\omega\  dy \label{eq:a}
\end{equation}
where
\[
   f(\omega,y)\equiv\omega-\Omega-\sqrt{(\omega-\Omega)^2-c^2-2c\mbox{Re}(a)\cos{y}-|a|^2\cos^2{y}}
\]
Taking the real and imaginary parts of~(\ref{eq:c}) and~(\ref{eq:a}) we obtain four real equations for the four real unknowns
$c,\mbox{Re}(a),\mbox{Im}(a)$ and $\Omega$.

As in Sec.~\ref{sec:model}, suppose that
\begin{eqnarray}
    g(\omega) & = & \frac{D/\pi}{\omega^2+D^2}=\frac{1}{2\pi i}\left[\frac{1}{\omega-iD}-\frac{1}{\omega+iD}\right]  
\end{eqnarray}
i.e.~the $\omega_i$ are from a distribution centred at zero with half-width-at-half-maximum $D$.
(There is no loss of generality by assuming that the distribution is centred at zero, since if this wasn't so, the effect 
would just be to add
a constant to $\Omega$.) Then for any function $F(\omega)$ analytic in the lower half of the complex $\omega$ plane,
\[
   \int_{-\infty}^{\infty}F(\omega)g(\omega)d\omega=F(-iD)
\]
and thus~(\ref{eq:c}) and~(\ref{eq:a}) become
\begin{equation}
    c=\frac{-e^{i\beta}}{2\pi}\int_0^{2\pi} \frac{\Omega+iD+\sqrt{(\Omega+iD)^2-c^2-2c\mbox{Re}(a)\cos{y}-|a|^2\cos^2{y}}}{c+\bar{a}\cos{y}}\ dy \label{eq:c1}
\end{equation}
and
\begin{equation}
   a=\frac{-Ae^{i\beta}}{2\pi}\int_0^{2\pi} \frac{\left(\Omega+iD+\sqrt{(\Omega+iD)^2-c^2-2c\mbox{Re}(a)\cos{y}-|a|^2\cos^2{y}}\right)\cos{y}}{c+\bar{a}\cos{y}}\ dy \label{eq:a1}
\end{equation}
Note that by setting $D=0$ in~(\ref{eq:c1}) and~(\ref{eq:a1})
we recover the results of Abrams and Strogatz~\cite{abrstr06}.

\subsection{Results}
\begin{figure}
\begin{center}
\includegraphics[width=12cm]{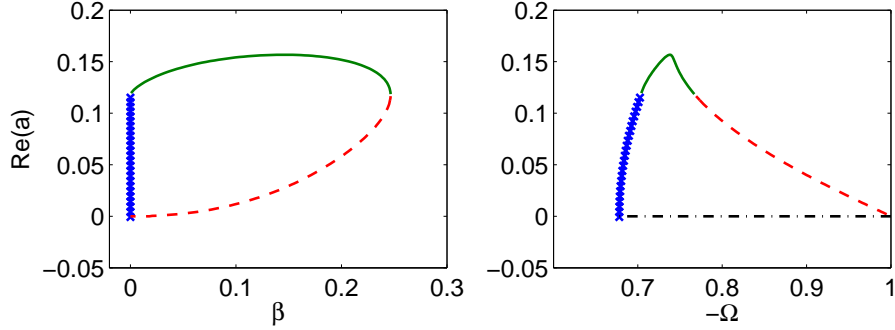}
\caption{Solutions of~(\ref{eq:c1}) and~(\ref{eq:a1}) when $D=0$. Left: Re($a$) versus $\beta$. Right: Re($a$) versus $-\Omega$, for $A=0.95$.}
\label{fig:iden}
\end{center}
\end{figure}

We now show results of following solutions of~(\ref{eq:c1}) and~(\ref{eq:a1}), using $D$ and $\beta$
as bifurcation parameters.
It is known for identical oscillators (i.e.~$D=0$) that for fixed $A>0$ chimerae exist for a 
range $0<\beta\leq \beta^\ast$,
and that $\beta^\ast$ is an increasing function of $A$ (see Fig.~8 in~\cite{abrstr06}). Here we set $A=0.95$.
Results for $D=0$
are shown in Fig.~\ref{fig:iden}; four types of solution are shown.

Blue crosses indicate the modulated drift state which occurs for $\beta=0,\mbox{Im}(a)=0$. In this state none of the oscillators
have synchronised with one another. The green solid line indicates the stable chimera, for which some of the
oscillators are synchronised with one another while the remainder drift. The red dashed curve is the unstable chimera.
The black dash-dot line represents the uniform drift state for which $\beta=a=0$. Along this line (which
has collapsed to a point in the left panel of Fig.~\ref{fig:iden}) all oscillators
are synchronised, $\theta_i=\theta_j, \forall i,j$. If $A$ was decreased, the saddle-node bifurcation seen in 
the left panel of Fig.~\ref{fig:iden}
would move to a lower value of $\beta$. (Note that stability of solutions is inferred, 
as it was by Abrams and Strogatz~\cite{abrstr06}.
All we have in~(\ref{eq:c1}) and~(\ref{eq:a1}) are algebraic equations governing the steady states, with no dynamics.)

\begin{figure}
\begin{center}
\includegraphics[width=12cm]{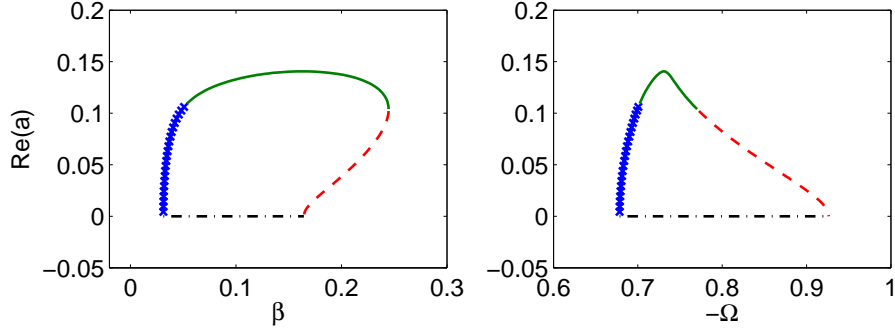}
\caption{Solutions of~(\ref{eq:c1}) and~(\ref{eq:a1}) when $D=0.01$. 
Left: Re($a$) versus $\beta$. Right: Re($a$) versus $-\Omega$, for $A=0.95$.}
\label{fig:noniden}
\end{center}
\end{figure}

A similar picture when $D=0.01$ is shown in Fig.~\ref{fig:noniden}, with the same conventions. We see that
the modulated drift state (which no longer has $\mbox{Im}(a)=0$) has moved away from $\beta=0$, as
has the uniform drift state ($a=0$).
Although we have not indicated it in Fig.~\ref{fig:iden}, for $D=0$ and $0<\beta$, the synchronised state (with $a=0$) is stable.
However for $0<D$ there is now a range of $\beta$ (approximately $0.03<\beta<0.165$ when $D=0.01$)
for which the synchronised state is unstable. (This is the extent of the dash-dotted line in the left panel of
Fig.~\ref{fig:noniden}.) For $\beta$ outside this range, i.e.~for $\beta$ small enough or large enough, the
synchronised state remains stable. 

Comparing the left panels of Fig.~\ref{fig:iden} and Fig.~\ref{fig:noniden} we see that the case of 
identical oscillators ($D=0$) is degenerate in the sense that both the modulated and uniform drift states
are ``hidden'' at $\beta=0$, but both occur over finite intervals of $\beta$ when $0<D$.

\begin{figure}
\begin{center}
\includegraphics[width=10cm]{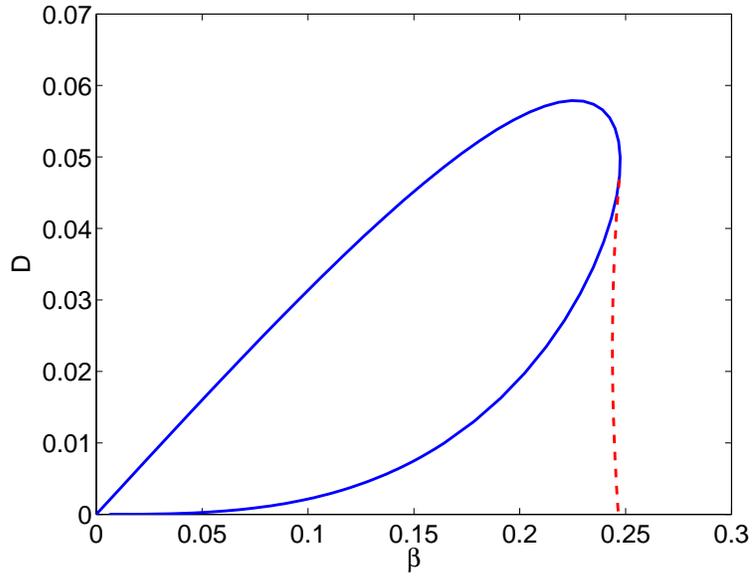}
\caption{Curves of pitchfork (solid) and saddle-node (dashed) bifurcations of solutions of~(\ref{eq:c1}) and~(\ref{eq:a1}).
Fig.~\ref{fig:noniden} corresponds to a horizontal ``slice'' through this figure at $D=0.01$. $A=0.95$.}
\label{fig:Dbeta}
\end{center}
\end{figure}

Fig.~\ref{fig:Dbeta} shows the results of following the pitchfork and saddle-node bifurcations seen in 
Fig.~\ref{fig:noniden} as $D$ is varied. Note that the two pitchfork bifurcations emanate from
$(D,\beta)=(0,0)$. The rightmost pitchfork bifurcation changes from sub- to super-critical
at the termination of the curve of saddle-node bifurcations. For approximately $0.047<D<0.058$ there is no bistability;
instead there is a range of $\beta$ values for which only the chimera state is stable. Outside this range only the
synchronous state is stable.
For $D$ greater than about 0.058 there do not exist
any chimera states. 

\begin{figure}
\begin{center}
\includegraphics[width=12cm]{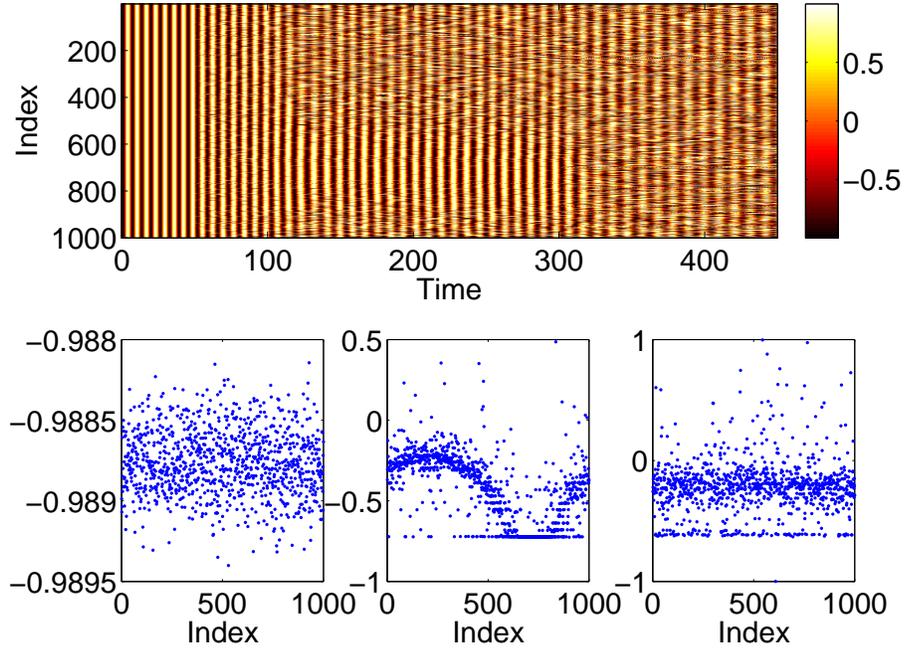}
\caption{Results of a simulation of~(\ref{eq:nettheta}) for which $D$ is switched from 0 to 0.02 at $t=50$,
and then increased to 0.06 at $t=300$. Top: $\sin{\phi_i}$. Bottom row: average of $d\phi_i/dt$ as a function of
index $i$ over the time intervals $[0,50]$ (left), $[150,300]$ (middle) and $[300,450]$ (right).
$N=1000,\beta=0.15,A=0.95$.}
\label{fig:demo}
\end{center}
\end{figure}

From Fig.~\ref{fig:Dbeta} we see that with $\beta$ small and fixed,
increasing $D$ first destabilises and then restabilises the synchronous state. This is the same behaviour
as observed in Fig.~\ref{fig:symm} for the model~(\ref{eq:dtheta}), and is demonstrated in Fig.~\ref{fig:demo}
where we fix $\beta=0.15$ and successively increase $D$ from 0 to 0.02 to 0.06. For $D=0$ the synchronised state
is stable (bottom left panel).
At $t\approx 100$ a chimera forms, with the center of the unsynchronised cluster at $i\approx 200$ and the 
center of the synchronised cluster at $i\approx 700$ (bottom middle panel). Once $D$ is increased above the 
upper blue curve in Fig.~\ref{fig:Dbeta} only the (noisy) synchronised state is stable (bottom right panel).

\subsection{Generalisations}
As shown in Sec.~\ref{sec:another},
if 
\[
    g(\omega)=\frac{\sqrt{2}D^3}{\pi}\left(\frac{1}{\omega^4+D^4}\right)
\]
we could repeat the analysis of~(\ref{eq:dphinet}), obtaining equations similar to, though more complicated than,~(\ref{eq:c1})
and~(\ref{eq:a1}), still with the unknowns $c,a$ and $\Omega$. Indeed, for any
distribution $g$ the double integrals~(\ref{eq:c}) and~(\ref{eq:a}) could be evaluated numerically.

The form of~(\ref{eq:Rtrig}) is a direct result of choosing $G$ to have one Fourier mode. If more modes were
used in $G$ (if, for example, we were approximating a given coupling function with a finite Fourier series), 
eqn.~(\ref{eq:Rtrig}) would have more terms and thus more coefficients to be found.

\subsection{Chimera states and ``bumps''}
Chimera states as studied in this section are very similar to ``bump'' states which
have been studied in computational neuroscience
modelling~\cite{lai06,lai01}. The main difference between bump states in neural models and the chimera
states studied here is that in neural models, the uncoupled unit is a model neuron which --- as an input current
is increased --- starts to fire periodically once the current has passed a threshold~\cite{lai01}, 
whereas in the chimera states
studied here the uncoupled unit is a phase oscillator with uniform angular velocity. In neural models, a bump
is a self-consistent state for which some neurons receive subthreshold input (and are thus quiescent)
while other receive superthreshold input (and are thus firing). It is the coupling of phase oscillators through
a sinusoidal function of the phase itself (as in~(\ref{eq:ptheta})) which allows some oscillators to lock and
rotate at a uniform frequency (which can be set to zero by moving to a rotating coordinate frame), 
while others drift, and thus a chimera can form.

To further emphasise the similarity, compare Fig.~4 in~\cite{lai01} with the insets in Fig.~12 in~\cite{abrstr06},
and with the middle plot in the bottom row of Fig.~\ref{fig:demo} (keeping in mind that this is a disordered system).
The bifurcations of bumps are also very similar to those of chimerae on a ring --- they both typically appear 
as unstable states bifurcating from a spatially-uniform state; compare Fig.~10 
in~\cite{lai06} with Fig.~12 in~\cite{abrstr06}.

\section{Summary}
We have considered the effects of heterogeneity in the intrinsic frequencies of oscillators on chimera states
in Kuramoto-type networks of coupled phase oscillators. Previous authors had only considered these 
states in networks of identical
oscillators~\cite{abrmir08,abrstr04,abrstr06,kurbat02,omemai08,setsen08}. By assuming a Lorentzian
distribution of intrinsic frequencies we have generalised the results of Abrams et al.~\cite{abrmir08} and
Abrams and Strogatz~\cite{abrstr04,abrstr06}, obtaining similar equations to them, but with an extra parameter,
viz.~the width of the Lorentzian distribution.
All of our results show that chimerae are robust --- within limits --- to heterogeneity in their intrinsic frequencies,
and we have shown some of the interesting bifurcations that can be induced by such heterogeneity.

Importantly, in light of the recent results of Pikovsky and Rosenblum~\cite{pikros08} regarding the 
validity of the Ott-Antonsen ansatz~(\ref{eq:ans}) used in this paper, our numerical results 
in Sec.~\ref{sec:results} support the observation
by Martens et al.~\cite{marbar08} that this ansatz can be used to study all attractors of a Kuramoto-type system
whenever the oscillators have randomly distributed frequencies.

The results presented here rely on the form of the equations studied. In particular, the results in
Secs.~\ref{sec:model}-\ref{sec:gen} rely on the remarkable recent results of Ott and Antonsen~\cite{ottant08}
showing that the infinite network can be exactly described by a finite number of ODEs, although not necessarily
completely~\cite{pikros08}.
Similarly, the analysis in Sec.~\ref{sec:ring} depended on the form of the coupling in~(\ref{eq:nettheta}),
through a trigonometric function of phase differences. The challenge remains to discover similar
results for oscillators not described by a single variable, and not coupled in this way.

\vspace{5mm}
\noindent
{\bf Acknowledgements:} I thank the referees for their very helpful comments.


\end{document}